%% file: main.tex
\PassOptionsToPackage{hyphens}{url}
\documentclass[sigplan,nonacm]{acmart}

\usepackage{multirow}
\usepackage{xspace}
\usepackage{listings}
\usepackage{pifont}
\usepackage{algorithm}
\usepackage[noend]{algpseudocode}

\definecolor{lstbg}{gray}{0.93}
\input{metadata.tex}
\let\acmparagraph\paragraph
\input{macros.tex}

\let\paragraph\acmparagraph

\renewcommand{\systemname}{TauLang}
\renewcommand{\papertitle}{A Language and Compiler for Space-Time Programming on Statically Scheduled Accelerators}

\begin{document}

\title[\systemname: \papertitle]{\systemname: \papertitle}

\author[Xuxiao Yang, Junchi Wu, Yuqi Peng, Feiyang Wu, Zhuoran Li, Wei Kong, Shuo Tao, Dongyi Fan, Yiming Zhang, Keqin Luo, Xingrui Gao, Hao Zhang, Xuehai Qian, and Youwei Zhuo]{%
  Xuxiao Yang\textsuperscript{1},
  Junchi Wu\textsuperscript{1},
  Yuqi Peng\textsuperscript{2},
  Feiyang Wu\textsuperscript{1},
  Zhuoran Li\textsuperscript{1},
  Wei Kong\textsuperscript{2},
  Shuo Tao\textsuperscript{2},
  Dongyi Fan\textsuperscript{2},
  Yiming Zhang\textsuperscript{2},
  Keqin Luo\textsuperscript{2},
  Xingrui Gao\textsuperscript{2},
  Hao Zhang \textsuperscript{2},
  Xuehai Qian\textsuperscript{2},
  Youwei Zhuo\textsuperscript{1}%
}

\affiliation[obeypunctuation=true]{%
  \institution{\textsuperscript{1}School of Integrated Circuits, Peking University}
  \city{Beijing}
  \country{China}
}

\affiliation[obeypunctuation=true]{%
  \institution{\textsuperscript{2}Tsinghua University}
  \city{Beijing}
  \country{China}
}

\affiliation[obeypunctuation=true]{%
  \institution{%
    \href{mailto:valinor@stu.pku.edu.cn,2501111908@stu.pku.edu.cn}{%
      \{valinor, 2501111908\}@stu.pku.edu.cn},
    \href{mailto:pyq24@mails.tsinghua.edu.cn}{%
      pyq24@mails.tsinghua.edu.cn},
    \href{mailto:fywu25@stu.pku.edu.cn,zhuoranli@stu.pku.edu.cn}{%
      \{fywu25, zhuoranli\}@stu.pku.edu.cn},
    \href{mailto:kongw24@mails.tsinghua.edu.cn,taos23@mails.tsinghua.edu.cn,fdy24@mails.tsinghua.edu.cn,zym24@mails.tsinghua.edu.cn,luokq24@mails.tsinghua.edu.cn,gxr24@mails.tsinghua.edu.cn,hao-zhang@mail.tsinghua.edu.cn}{%
      \{kongw24, taos23, fdy24, zym24, luokq24, gxr24, hao-zhang\}@mails.tsinghua.edu.cn},
    \href{mailto:xuehaiq@tsinghua.edu.cn}{xuehaiq@tsinghua.edu.cn},
    \href{mailto:youwei@pku.edu.cn}{youwei@pku.edu.cn}%
  }
  \city{}
  \country{}
}

\begin{abstract}
\input{sections/abstract.tex}
\end{abstract}

\maketitle

\input{content.tex}

\bibliographystyle{ACM-Reference-Format}
\bibliography{references}

\end{document}

%% file: metadata.tex
\newcommand{\papertitle}{Time-First Programming for Statically Scheduled Accelerators}

%% file: macros.tex
\renewcommand{\paragraph}[1]{\par\noindent\textbf{#1}}
\newcommand{\systemname}{TauVM}

\newcommand{\sys}{\systemname\xspace}
\newcommand{\irname}{TauIR}
\newcommand{\sysir}{\irname\xspace}

\newcommand{\code}[1]{\texttt{#1}}
\newcommand{\tissue}{\ensuremath{t_{\mathrm{issue}}}}
\newcommand{\mk}[1]{\ding{\numexpr 201+#1\relax}}

%% file: sections/abstract.tex
Statically scheduled accelerators such as Groq's Tensor Streaming Processor (TSP) achieve low-latency large language model inference by stripping dynamic synchronization hardware from the chip.
With no caches, arbiters, or hardware schedulers, the compiler must fix the exact slice and issue cycle of every operation, so \emph{when} an operation runs is a condition for correctness.
Existing kernel languages cannot target such machines because they rely on hardware to resolve cycle timing and instead encode schedules implicitly in code structure.

We present \sys, a kernel language and compiler that makes time a first-class citizen of the program.
The programmer schedules operations, loops, and composed kernels with parameterized explicit times,
so exploring schedules means changing parameter values rather than rewriting code.
\sys validates explicit times algebraically in \sysir, its timed intermediate representation.
It then solves tensor placement and cycle scheduling jointly in an integer linear program,
using explicit times to group connected operations into timing clusters that prune the search space.
On a cycle-accurate model of a TSP-like machine, joint search achieves a $1.40\times$ latency speedup across ten kernels
over a space-first flow.
A few explicit time pins keep latency within one cycle of the untimed schedules
and scale joint search to 1,636 GEMM blocks, whereas untimed search finds no schedule at 32.

%% file: content.tex
\input{sections/introduction.tex}
\input{sections/background-motivation.tex}
\input{sections/overview.tex}
\input{sections/language.tex}
\input{sections/timed-ir.tex}
\input{sections/space-time-scheduling.tex}
\input{sections/implementation.tex}
\input{sections/evaluation.tex}
\input{sections/related-work.tex}
\input{sections/conclusion.tex}

%% file: sections/introduction.tex
\section{Introduction}
\label{sec:introduction}

Statically scheduled accelerators, pioneered by Groq's Tensor Streaming Processor (TSP)~\cite{abts2020thinkfast,abts2022software}, have emerged as a compelling architecture for low-latency large language model inference.
Reflecting this demand for fast token generation, Nvidia recently acquired Groq's assets for \$20 billion~\cite{cnbc2025nvidiagroq} and introduced a rack-scale system built on the TSP execution model~\cite{nvidia2026lpxinside}.
To achieve such low, deterministic latency, this execution model strips out the dynamic control hardware of conventional accelerators, eliminating caches, local register files, interconnect arbiters, and hardware schedulers~\cite{singh2022virtuous}.

\begin{figure}[t]
\centering
\includegraphics[width=\columnwidth]{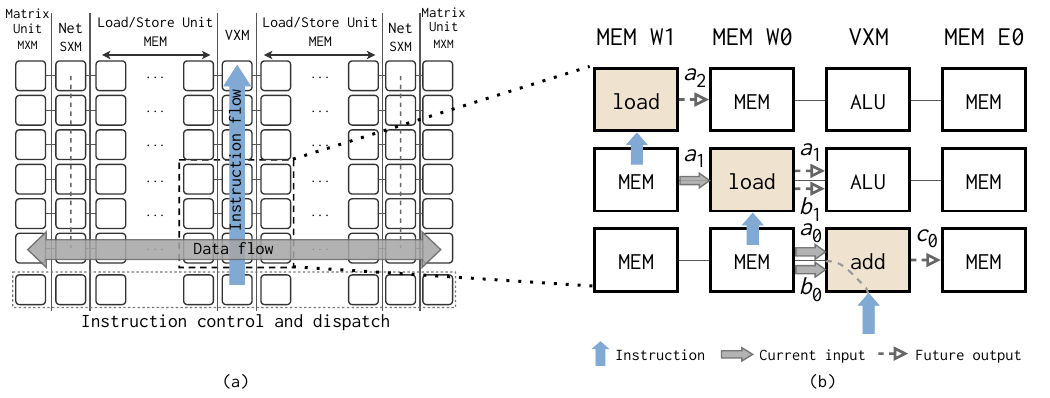}
\caption{\textbf{(a)} Instructions flow vertically while data streams horizontally across slices.
\textbf{(b)} Staggered \code{load}s make data $a_0, b_0$ and the \code{add} instruction meet at \code{VXM} in the same cycle.}
\label{fig:arch}
\vspace{-7mm}
\end{figure}

However, the simplicity of the hardware makes compilation harder.
With no stall signals or runtime handshakes, instructions flow vertically up each slice while data streams move horizontally across functional units at a fixed rate of one slice per cycle, without waiting (Figure~\ref{fig:arch}a).
Thus, the compiler must statically decide the exact slice and issue cycle of every operation so that each instruction and its input streams meet at the target unit in the exact same cycle (Figure~\ref{fig:arch}b), while avoiding conflicts on shared stream register (SR) tracks and memory banks.
Because the hardware detects no timing or resource violations, missing a rendezvous by even one cycle or slice silently produces wrong results, which makes \emph{when} an operation runs a condition for correctness rather than a performance parameter.
Moreover, a single chip contains dozens of slices, so the compiler must coordinate these cycle-exact schedules across all slices and functional units simultaneously.

Existing approaches to this challenge fall between two extremes.
On one hand, expert engineers can hand-craft low-level instruction schedules~\cite{abts2022software,singh2022virtuous}, but doing so demands painstaking cycle-level control and months of effort.
For example, after Groq initially brought up Llama-2 70B at 100 tokens/s, its engineers spent three months optimizing low-level schedules on the same hardware to reach 300 tokens/s~\cite{groq2023llama300}.
On the other hand, a compiler that schedules high-level code automatically faces a vast joint space of slice placements and issue cycles.
Fixing placement first shrinks this space but yields slower schedules,
whereas in our experiments, an unconstrained solver finds no schedule for a 32-block GEMM within the time budget.
A natural middle ground is a kernel language that lets the programmer express high-level scheduling decisions while the compiler checks and completes the low-level schedule.

However, neither GPU kernel languages~\cite{tillet2019triton,wang2026tilelang} nor spatial compilers~\cite{li2026tileloom,mei2003dresc,liu2019cgrasurvey} can serve as this middle ground, because they assume that hardware synchronizes timing at runtime.
Even compilers for spatial meshes and coarse-grained reconfigurable arrays (CGRAs) that plan placement and modulo cycles statically~\cite{li2026tileloom,mei2003dresc,liu2019cgrasurvey} rely on scratchpads, FIFOs, and router flow control to buffer early arrivals~\cite{prabhakar2017plasticine,gobieski2022riptide}, so a one-cycle mismatch costs only throughput.
Under this assumption, existing kernel languages are \emph{time-blind} and instead express schedules indirectly through code structure, such as pipeline stage counts and explicit barriers (Figure~\ref{fig:motivation-tilelang}).
This structural approach neither expresses cycle-exact rendezvous nor separates the schedule from the abstract syntax tree (AST), so the programmer must rewrite the kernel to change its schedule.

Our key insight is to decouple a program's schedule from its code structure by making \emph{time a first-class citizen} of the kernel language.
Because cycle timing determines correctness on a statically scheduled accelerator, the language should let the programmer express cycle times directly and the compiler check them statically (Figure~\ref{fig:motivation-tauvm}).
Programmers write \emph{explicit times} as affine cycle expressions over loop indices and kernel parameters,
and \emph{pinning} a kernel call to an invocation time offsets every explicit time in the callee.
Exploring schedules across a composed program thus requires only changing parameter values rather than rewriting the kernel.
Pins also keep the compiler's joint search over the remaining schedule tractable.
Because streams advance one slice per cycle,
pinning the cycle gap between a stream's producer and consumer fixes the hop count between their slices, which prunes the memory placements that the solver must explore.

We present \sys, a kernel language and compiler for statically scheduled accelerators.
\sys extends a tile-style Python DSL with explicit times relative to \emph{symbolic times},
which mark the start of the current kernel (\code{tau.T}) or the start or end of an operation, loop, or call (\code{tau.start}, \code{tau.end}).
A pin (\code{.at}) fixes the issue time of an operation or the invocation time of a call,
while \code{tau.domain} maps each loop iteration to its start cycle.
The compiler lowers explicit times into \sysir attributes in MLIR~\cite{lattner2021mlir} and runs \emph{affine time validation} to catch timing conflicts algebraically before search.
It then solves spatial placement and cycle scheduling jointly as an integer linear program (ILP), using stream dependences and explicit times to group connected operations into \emph{timing clusters} that prune the search.

We evaluate \sys on a TSP-like machine using a cycle-accurate simulator validated against our SystemVerilog RTL implementation.
Across ten kernels from BLAS, PolyBench, ResNet, and Transformer layers at three sizes each, joint search achieves a $1.40\times$ geometric-mean latency speedup (up to $2.22\times$)
over the space-first flow, which fixes placement before scheduling greedily.
Adding one to three pins per kernel matches untimed latency within one cycle
and cuts compile time by $1.72\times$.
Pins also scale joint search to 1,636 GEMM blocks and 128 residual blocks,
whereas untimed search finds no schedule at 32 and 8.
Sweeping two schedule parameters of a flow-matching kernel yields eight Pareto-optimal schedules among those compiled,
which trade up to 237 cycles of first-item latency for 498 cycles of whole-output time.

This paper makes three contributions.
\begin{itemize}
  \item \textbf{A time-first kernel language.} We design \sys, to our knowledge the first kernel language that treats cycle time as a first-class, parameterized, and composable construct (\S\ref{sec:kernels}).
  \item \textbf{A timed intermediate representation.} We define \sysir, an IR that carries symbolic timing attributes and validates them algebraically before scheduling (\S\ref{sec:timed-ir}).
  \item \textbf{Time-guided joint space-time scheduling.} We formulate placement and scheduling as a joint ILP that uses explicit times to group operations into timing clusters and prune the search space (\S\ref{sec:scheduling}).
\end{itemize}

%% file: sections/background-motivation.tex
\section{Background and Motivating Examples}
\label{sec:background}

\subsection{The TSP Execution Model}
\label{sec:machine}

Groq's TSP~\cite{abts2020thinkfast,abts2022software} is the most widely deployed statically scheduled accelerator, so we use it as our target architecture.
On the row of slices in Figure~\ref{fig:arch}a, instructions step vertically up each slice across its lanes while values travel east or west on SR tracks at one slice per cycle.
Memory slices (\code{MEM}) store tensors in SRAM banks, vector (\code{VXM}) and switch (\code{SXM}) slices perform arithmetic and lane permutations, and matrix slices (\code{MXM}) hold weights across two planes so one plane can install weights while the other computes.
Because execution and stream propagation are deterministic, a consumer sees a producer's output at a fixed cycle.
An operation samples its operands $d_{\mathrm{skew}}$ cycles after it issues and writes its result to an SR $d_{\mathrm{func}}$ cycles after it issues.
Let $\delta$ be the hop count from producer $P$ to consumer $C$.
$C$ receives $P$'s value if and only if
\begin{equation}
\tissue(P) + d_{\mathrm{func}}(P) + 1 + \delta \;=\; \tissue(C) + d_{\mathrm{skew}}(C).
\label{eq:rendezvous}
\end{equation}
Figure~\ref{fig:arch}b illustrates this rendezvous for $c = a + b$, where \code{MEM W1} is one slice farther from \code{VXM} than \code{MEM W0} and therefore issues its \code{load} of $a$ one cycle earlier so that $a_0$ and $b_0$ reach \code{add} on \code{VXM} in the same cycle.
Furthermore, at most one producer may write a given SR track at any slice and cycle, and each memory bank serves at most one access per cycle.
The hardware detects neither violation and silently produces wrong values instead.

\subsection{Schedules in Existing Kernel Languages}
\label{sec:motivation}

\begin{figure}[t]
\centering
\includegraphics[width=\columnwidth]{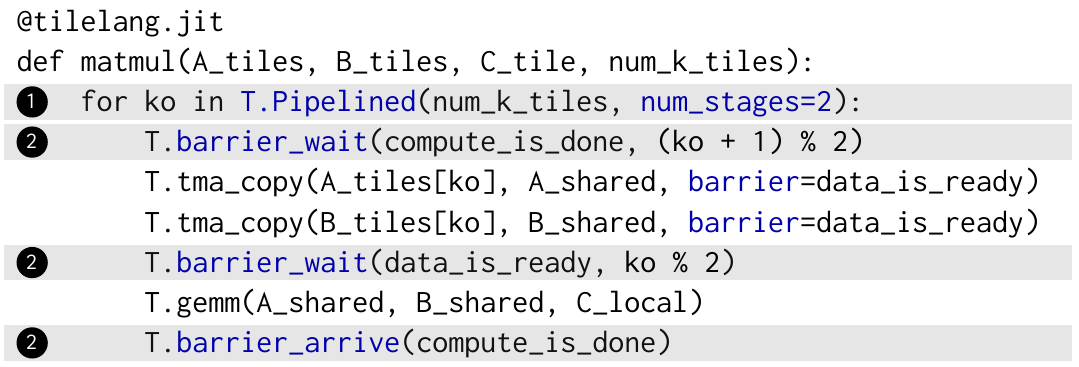}
\caption{A pipelined GEMM in TileLang.
The schedule is encoded in code structure via stage counts (\mk{1}) and runtime barriers (\mk{2}).}
\label{fig:motivation-tilelang}
\vspace{-6mm}
\end{figure}

Because the best schedule of a kernel varies across workloads, the programmer needs fine-grained control over when each operation executes.
On a statically scheduled machine, this schedule also determines correctness, as every operand must arrive at the exact slice and cycle its instruction expects.
We examine how TileLang~\cite{wang2026tilelang} expresses the schedule of a GEMM and why that schedule does not carry over to the TSP.

TileLang describes a kernel as an iteration domain whose body applies tile operators such as load, store, and matrix multiply to whole tiles~\cite{wang2026tilelang}.
In the pipelined GEMM in Figure~\ref{fig:motivation-tilelang}, \code{T.Pipelined} specifies a stage count~\mk{1} to overlap iterations through multi-buffered copies, while runtime barriers synchronize stages~\mk{2} as each multiply waits for its copies and each pair of copies waits for the previous multiply.\footnote{The code listings in this section omit type annotations, explicit casts, and routine implementation details.}

Thus, the loop runs at whatever pace the copies and barriers deliver at runtime.
Both constructs rely on hardware synchronization, where copies signal completion and barriers stall waiting warps.
On the TSP, nothing waits, because a consumer receives a value only in the cycle required by Equation~\ref{eq:rendezvous}.

As this GEMM shows, existing kernel languages share two limitations.
\textbf{First}, they cannot express cycle timing, so the programmer cannot specify even the initiation interval of a loop.
\textbf{Second}, the control they do expose is coarse and written in the structure of the code, so another schedule is another program.
For instance, alongside the kernel in Figure~\ref{fig:motivation-tilelang}, TileLang provides a separate program to swap which warp group copies tiles and which performs matrix multiplication.
Both limitations arise because time is not a first-class citizen of these languages.
\sys addresses both by keeping the loops and tile operators of a tile-style language and adding \emph{explicit times}, cycle counts written over ordinary kernel parameters, so another schedule is simply another set of parameter values on the same program.
Sections~\ref{sec:examples}--\ref{sec:examples-attention} preview through concrete examples how explicit times parameterize schedules, enable static timing checks, and prune placement and scheduling.

\subsection{Explicit Times in a Tiled GEMM}
\label{sec:examples}

\begin{figure}[t]
\centering
\includegraphics[width=\columnwidth]{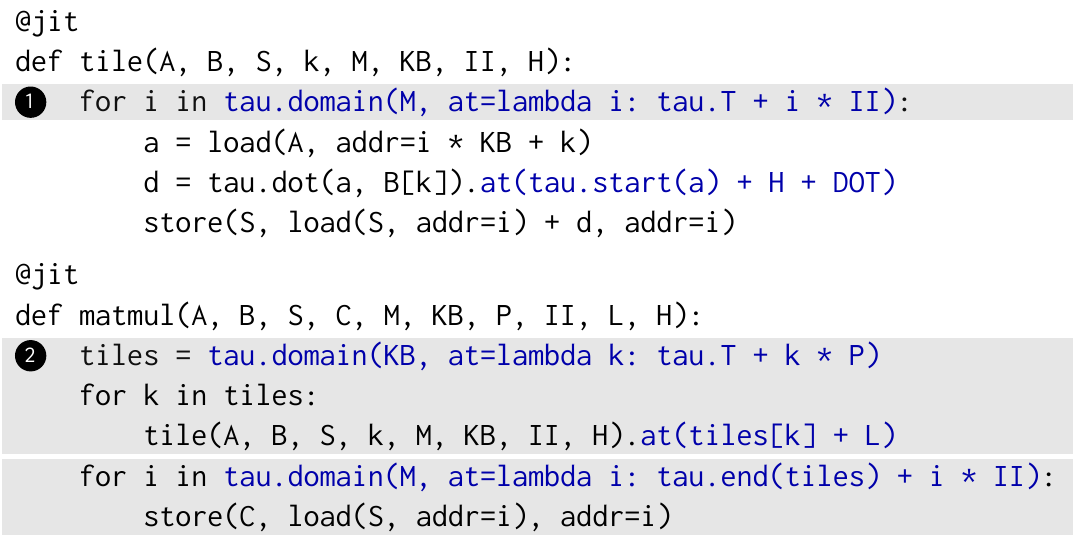}
\caption{A tiled GEMM in \sys.
Explicit times start rows every $\mathit{II}$ cycles (\mk{1}) and tiles every $P$ cycles (\mk{2}).}
\label{fig:motivation-tauvm}
\end{figure}

\begin{figure}[t]
\centering
\includegraphics[width=\columnwidth]{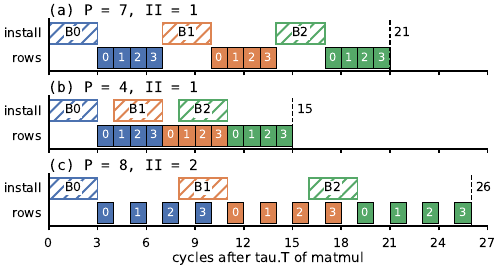}
\caption{Three schedules of the GEMM in Figure~\ref{fig:motivation-tauvm}, set by the tile stride $P$ and the row interval $\mathit{II}$.}
\label{fig:schedules}
\end{figure}

Figure~\ref{fig:motivation-tauvm} writes the same tiled GEMM in \sys, where \code{matmul} calls \code{tile} on each block along $K$ to install weights on an \code{MXM} plane and stream rows of $A$.
Rather than encoding the schedule in stage counts and barriers, \sys states loop strides directly in cycles with \code{tau.domain} and fixes selected stage offsets with \code{.at}.
Inside \code{tile}, the row domain (\mk{1}) starts a new row of $A$ every $\mathit{II}$ cycles, and \code{.at} pins the accumulation at $H + \mathit{DOT}$ cycles after \code{load(A)}, where $H$ is the hop count from \code{load(A)} to the matrix unit.
In \code{matmul}, the \code{tiles} domain (\mk{2}) starts a new block of $B$ every $P$ cycles and leaves $L$ cycles to install weights before calling \code{tile}.
The final loop drains the partial sums $S$ to $C$ once the last tile finishes.

Because $P$ and $\mathit{II}$ are ordinary kernel parameters, exploring schedules reduces to changing parameter values on the same code (Figure~\ref{fig:schedules}).
A large tile stride runs the three tiles sequentially, waiting for one tile's rows to finish before installing the next weight block (Figure~\ref{fig:schedules}a).
Shortening $P$ pipelines adjacent tiles across the two \code{MXM} planes so each weight install overlaps the previous tile's rows (Figure~\ref{fig:schedules}b), while raising $\mathit{II}$ spaces rows apart to free cycles for another kernel running alongside (Figure~\ref{fig:schedules}c).
Section~\ref{sec:case-study} sweeps two such parameters on a flow-matching kernel to trade first-item latency against whole-output time.

\subsection{Checking Explicit Times}
\label{sec:examples-check}

Because explicit times directly control when operations issue, an invalid schedule can violate data dependences.
Suppose the programmer extends the GEMM across column blocks of $B$ and starts the final loop early to overlap it with the tile walk.
If the final loop drains columns faster than the walk finishes accumulating them, later columns read elements of $S$ before their last write.
Whereas a GPU barrier would stall such a read until the write completes, the TSP reads the stale partial sum from SRAM and silently corrupts $C$.
\sys rejects the program before scheduling with a read-before-write error on $S$ that cites the offending source lines, and accepts it once each column of the final loop starts after its tile walk finishes (\S\ref{sec:ir-validation}).

\subsection{Pinning One Stage in Attention}
\label{sec:examples-attention}

\begin{figure}[t]
\centering
\includegraphics[width=\columnwidth]{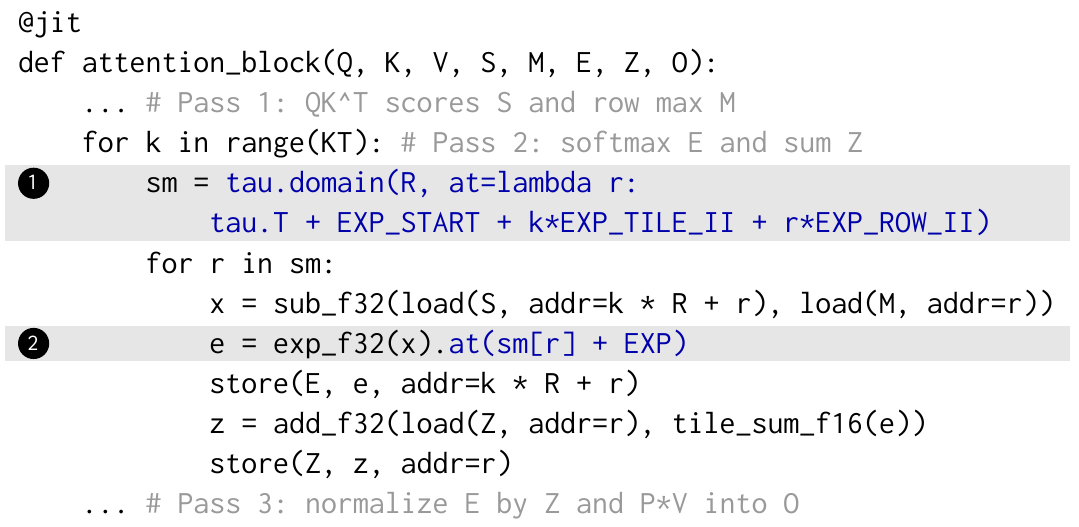}
\caption{The softmax pass of a three-pass tiled attention kernel in \sys.
The pass starts \code{EXP\_START} cycles after \code{tau.T}, its key tiles start \code{EXP\_TILE\_II} cycles apart and its rows \code{EXP\_ROW\_II} apart (\mk{1}), and \code{exp\_f32} issues \code{EXP} cycles after its row starts (\mk{2}).}
\label{fig:attention}
\end{figure}

Pins are optional, so in a loop body where multiple streams reconverge, the programmer pins only the stages of interest.
Figure~\ref{fig:attention} shows such a body in the softmax pass of a three-pass attention kernel, after pass~1 has computed the scores $S$ and the row maxima $M$.
The pass iterates over the $\mathit{KT}$ key tiles in a plain \code{range} loop, while the inner \code{tau.domain} schedules both loops by incorporating the outer index $k$ into its time map (\mk{1}).
Each row computes $e = \exp(S - M)$, stores it to $E$, and adds its cross-lane sum \code{tile\_sum\_f16(e)} to the running denominator in $Z$.
By Equation~\ref{eq:rendezvous}, the scores and the row maximum must reach \code{sub\_f32} in the same cycle, as must the row sum and the denominator loaded from $Z$ at \code{add\_f32}.

The \code{.at} call (\mk{2}) pins \code{exp\_f32} \code{EXP} cycles after its row starts.
Because the time map fixes the row start, the pin fixes every cycle gap along the path from the loads of $S$ and $M$ through \code{sub\_f32} to \code{exp\_f32}.
Beyond fixing that timing, the pin turns the path into one timing cluster whose cycle gaps and hop counts the solver no longer searches, which is how pins cut compile time in \S\ref{sec:eval-written}.
The reduction tail after \code{exp\_f32} carries no explicit time, and the compiler places and schedules the row sum, the load of $Z$, and \code{add\_f32} so that their operands arrive in the same cycle.

%% file: sections/overview.tex
\section{Overview}
\label{sec:overview}

%

\begin{figure}[h!]
  \centering
  \includegraphics[width=\columnwidth]{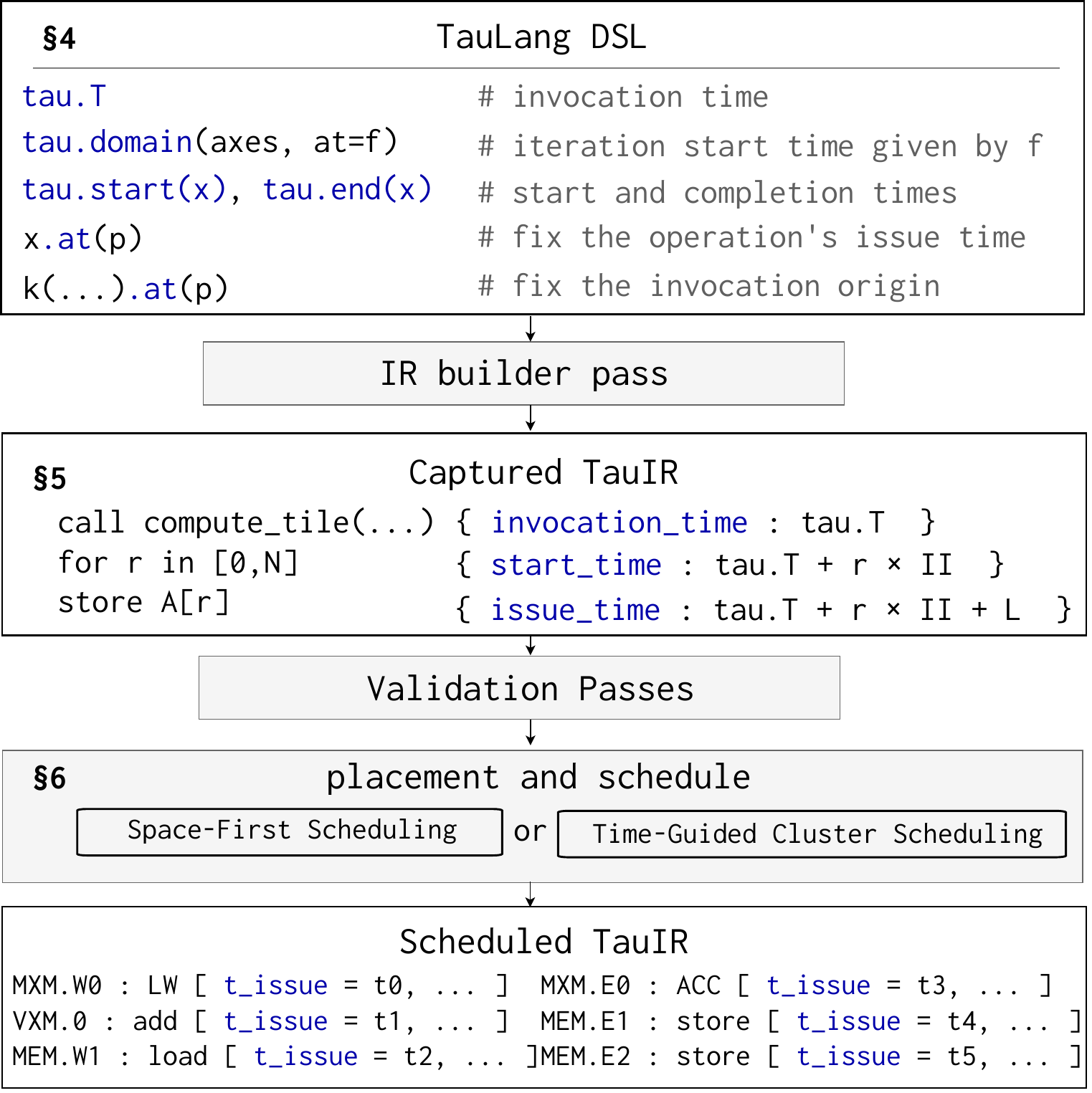}
  \caption{The \sys workflow, from the DSL to the scheduled \irname.
  Each explicit time remains an attribute of the call, loop, or operation it schedules
  until placement and scheduling assign slices and issue cycles.}
  \label{fig:overview}
\end{figure}

Figure~\ref{fig:overview} shows how \sys compiles a kernel written with explicit timing constructs (\S\ref{sec:kernels}) into one timestamped program per slice.
An IR builder pass first captures the kernel into \sysir (\S\ref{sec:timed-ir}).
As the captured \sysir box illustrates, each call carries an \code{invocation\_time}, each timed loop carries a \code{start\_time}, and each pinned operation carries an \code{issue\_time}.
Before any slice or cycle is assigned, validation passes check these attributes algebraically for timing errors that are independent of placement,
such as the overlapped final loop of \S\ref{sec:examples-check}.

Placement and scheduling then run in one of two flows (\S\ref{sec:scheduling}) to assign slices and issue cycles for all operations while satisfying Equation~\ref{eq:rendezvous} and every explicit time.
The cluster flow, our third contribution, formulates joint search as an ILP that decides where each tensor buffer resides in memory and when every operation issues.
To keep this search tractable, the flow first merges operations whose cycle gaps are fixed into timing clusters.
The space-first flow serves as a fast baseline that assigns slices using deterministic rules and then schedules issue cycles greedily.
Both flows emit the scheduled \sysir at the bottom of the figure, which lists each operation on its assigned slice with its issue cycle.
Code generation then assembles the scheduled \sysir into per-slice instruction streams (\S\ref{sec:implementation}).

%% file: sections/language.tex
\section{The \sys Language}
\label{sec:kernels}

As previewed in \S\ref{sec:examples} and \S\ref{sec:examples-attention}, \sys extends a tile-style Python DSL with explicit times, which the programmer specifies where timing matters while the compiler schedules the rest.
Every explicit time is an affine cycle offset from a symbolic \emph{root} over the enclosing loop indices and compile-time kernel parameters.
Symbolic times supply the roots, pins attach a time to an operation or a call, and iteration domains attach a time map to a loop.

\subsection{Symbolic Times}
\label{sec:kernels-symbolic}

Absolute cycle numbers from the start of the program would break modularity whenever a sub-kernel is reused or preceded by other stages.
Instead, every explicit time is an offset from one of three \emph{symbolic times}, \code{tau.T}, \code{tau.start}, and \code{tau.end}.
\code{tau.T} denotes the start of the current kernel invocation.
\code{tau.start(x)} and \code{tau.end(x)} return the symbolic time at which an operation, iteration domain, or kernel invocation \code{x} starts or ends.
A time measured from \code{tau.start(x)} or \code{tau.end(x)} constrains only the cycle distance from event \code{x} while leaving the exact cycle of \code{x} for the compiler to choose, as the accumulation pin and the final loop of Figure~\ref{fig:motivation-tauvm} do.

\subsection{Pinning a Time}
\label{sec:kernels-at}

A \emph{pin}, written with the \code{.at} method, fixes the time at which one operation or kernel invocation executes.
Written on a value, \code{x.at(p)} states that the operation producing $x$ issues at time $p$, while its result reaches its consumers after the delays of Equation~\ref{eq:rendezvous}.
The compiler chooses the issue cycle of an operation without \code{.at}, subject to the time map of an enclosing domain.
A pin's time expression may include architectural latency constants queried via \code{arch\_int}, such as \code{DOT} in the accumulation pin of Figure~\ref{fig:motivation-tauvm}.

Written on a call, \code{k(...).at(p)} sets the \code{tau.T} of callee $k$ to the caller's time $p$, which serves as the call's \emph{invocation time} and shifts every explicit time in the callee by $p$.
The \code{tau.T} of the entry kernel is the start of the program.
If a callee contains no explicit times, its body starts at the invocation time.

\subsection{Iteration Domains}
\label{sec:kernels-domain}

\code{tau.domain} schedules a loop by declaring an iteration space together with a \emph{time map} (\code{at=lambda ...}) from loop indices to the start of each iteration.
In polyhedral terms, a time map defines an affine schedule from the iteration space to cycle time~\cite{feautrier1992affine}, where the coefficient of a loop index sets the initiation interval along that axis in modulo scheduling~\cite{rau1994modulo}.
The earliest unpinned operation of each iteration initiates the iteration at the exact cycle specified by the time map, and every other unpinned operation in the loop body issues at or after that cycle.
An unpinned call takes its iteration start as its invocation time.
Binding a domain to a name, such as \code{tiles} in Figure~\ref{fig:motivation-tauvm}, lets the program refer to its iteration starts, so \code{tiles[k]} is the start of iteration $k$ and \code{tau.end(tiles)} is the end of the domain.
When a domain is written without \code{at}, it declares only the bounds and the compiler chooses the initiation interval of the loop.

Because a time map can reference any enclosing loop index, a domain nested inside an untimed \code{range} loop can incorporate the outer index into its schedule.
In the softmax pass of Figure~\ref{fig:attention}, the term \code{k*EXP\_TILE\_II} in the time map (\mk{1}) starts each key tile's domain \code{EXP\_TILE\_II} cycles after the previous one, which schedules the outer loop without a separate time map.


%% file: sections/timed-ir.tex
\section{\irname}
\label{sec:timed-ir}

The compiler lowers explicit times from the DSL into attributes of \sysir, the intermediate representation of \sys.
This section describes how \sysir composes and lowers these timing attributes onto a shared timeline and how affine time validation rejects inconsistent schedules before placement and scheduling.

\subsection{Timing Attributes}
\label{sec:ir-attributes}

\sysir represents each explicit time as one of three symbolic affine timing attributes in MLIR~\cite{lattner2021mlir} (Figure~\ref{fig:overview}).
A call carries an \code{invocation\_time}, a timed loop carries a \code{start\_time} map over its loop indices, and an operation carries an \code{issue\_time} \emph{pin}.
\sysir carries no unroll factor or pipeline stage count, so these attributes capture the entire programmer-specified schedule.

At the operation level, \sysir distinguishes \emph{explicit} and \emph{derived} pins, each measured from a symbolic root.
A pin the programmer writes with \code{.at} is an \emph{explicit pin}, which fixes the exact cycle (from \code{tau.T}) or cycle distance (from \code{tau.start} or \code{tau.end}) at which the operation issues.
Every unpinned operation inside a \code{tau.domain} instead carries a \emph{derived pin}, the time derived from the iteration start of its nearest enclosing timed domain, which the operation takes as a lower bound on its issue cycle.
Among these operations, only those with no intra-iteration operands, such as a load, can issue first, so \sysir marks them as candidates to initiate the iteration.
The earliest scheduled candidate initiates the iteration at the exact cycle of its derived pin.
Each iteration thus starts at the cycle specified by the time map without requiring the programmer to pin an operation by hand.
Validation and cluster partitioning rely on these operations for exact cycle gaps across iterations.
An unpinned operation outside every \code{tau.domain} has no derived pin and issues wherever its dependences and resources allow.

\sys lowers these timing attributes onto a single timeline in three stages while preserving their meaning.
First, it composes every call into the entry kernel by adding the call's \code{invocation\_time} to every time attribute of the callee, which places all times of the program on one shared timeline.
Second, for each unpinned operation inside a \code{tau.domain}, it walks outward to the nearest enclosing timed domain to attach its derived pin and marks the candidate operations that can initiate each iteration.
Finally, it unrolls the inner loops of every loop nest, while each loop that stays rolled takes the coefficient of its index in the pins as its initiation interval.
Each unrolled instance retains the source form of its time attribute, so any diagnostic on a lowered operation quotes the line the programmer wrote.

\subsection{Affine Time Validation}
\label{sec:ir-validation}

\begin{figure}[t]
\centering
\includegraphics[width=\columnwidth]{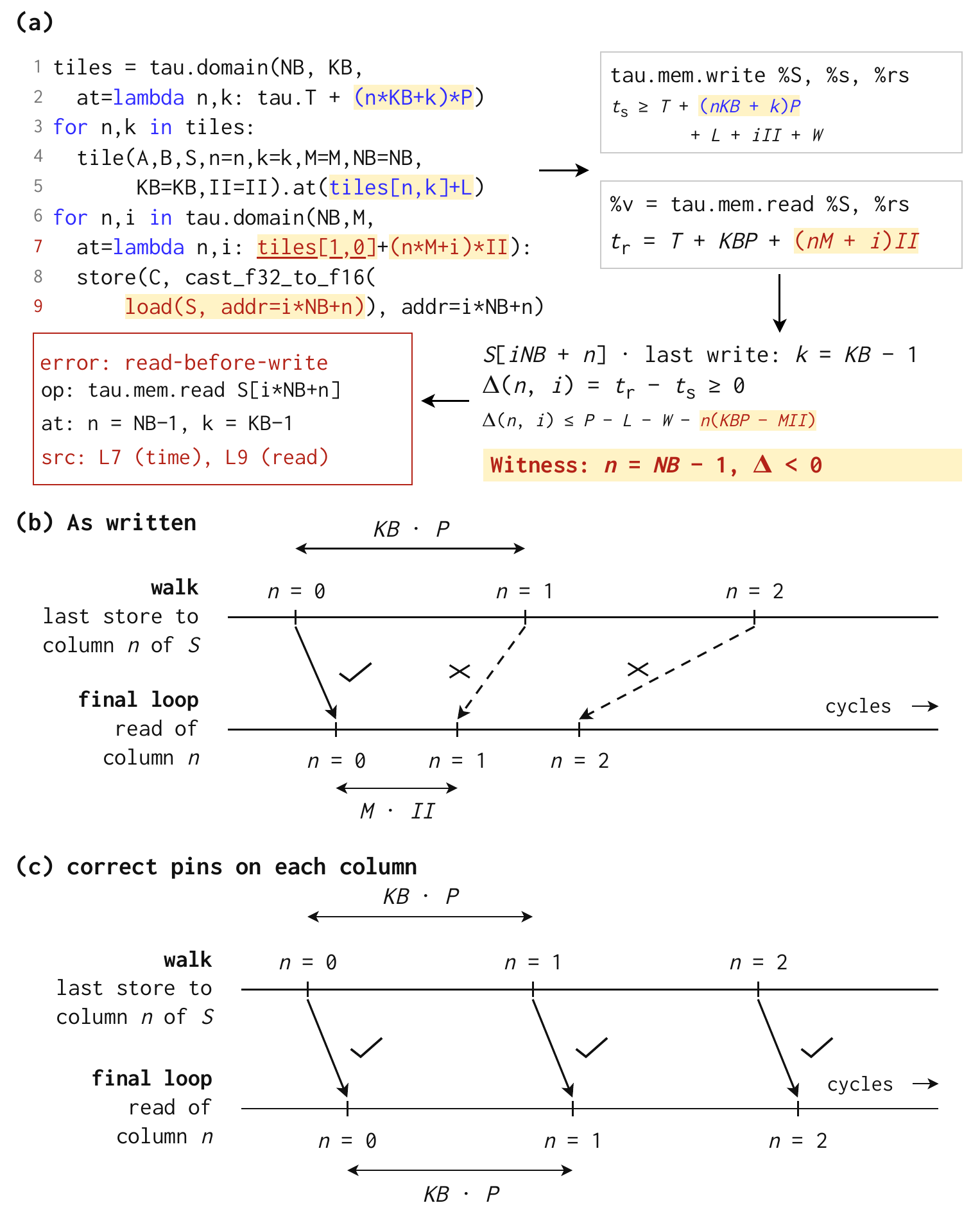}
\caption{Validation rejects a tiled GEMM whose final loop reads $S$ before the tile walk finishes writing it.
\textbf{(a)} The program, the composed times of the write and the read of $S$, and the diagnostic.
\textbf{(b)} The paces of the walk and of the final loop as written.
\textbf{(c)} The same paces once each column of the final loop starts after its tile walk.}
\label{fig:time-validation}
\end{figure}

As illustrated by the read-before-write bug in \S\ref{sec:examples-check}, explicit times may conflict with machine latencies or with one another.
Before placement and scheduling run, \sys checks the explicit times algebraically on \sysir against \emph{necessary timing conditions} that hold regardless of slice placement and leaves conflicts that depend on exact hop counts to placement and scheduling.
These conditions are independent of placement because the explicit times, the dependence graph, and the opcode latencies ($d_{\mathrm{func}}, d_{\mathrm{skew}}$) are already fixed in \sysir, whereas placement only varies hop counts within chip bounds.
Each check takes the difference of two times sharing a root and verifies that it satisfies these necessary bounds across all loop indices.
Validation rejects three error patterns, namely (1) a conflict among explicit times, such as a pin (\code{.at}) that places an operation before its iteration starts, (2) an unbalanced pipeline in which reconvergent data paths have different cycle lengths, and (3) a data-dependence violation in storage, such as reading before a write commits or overwriting a value before it is read, which silently produces wrong data.

Figure~\ref{fig:time-validation} traces the third pattern on the overlapped final loop of \S\ref{sec:examples-check}.
On the composed timeline, each read of $S$ in the final loop (at $t_r$, which initiates its iteration) must follow the last write to that element in tile $(n, \mathit{KB}-1)$ (at $t_s \ge$ row start $+ W$, where $W$ is the minimum load-to-store latency through a row).
This requires the upper bound $P - L - W - n(\mathit{KB} \cdot P - M \cdot \mathit{II})$ of $t_r - t_s$ to remain non-negative.
Because the final loop drains columns faster than the walk finishes them (Figure~\ref{fig:time-validation}b), $t_r - t_s < 0$ from $n = 1$ through $n = \mathit{NB}-1$.
Validation catches this negative slack and reports the offending source lines (Figure~\ref{fig:time-validation}a).
Starting column $n$ at \code{tiles[n + 1, 0]} cancels the term in $n$ and leaves the bound $P - L - W$, so validation accepts the program when $P \ge L + W$ (Figure~\ref{fig:time-validation}c).

%% file: sections/space-time-scheduling.tex
\section{Space-Time Scheduling}
\label{sec:scheduling}

Once affine time validation accepts the explicit times, the compiler must still decide where and when every operation runs.
Placement assigns each operation to a slice, and scheduling assigns its issue cycle.
Both choices must satisfy Equation~\ref{eq:rendezvous} as well as the per-cycle bandwidth limits of SR tracks and memory banks.
Because values on a stream advance one slice per cycle without waiting, placement and scheduling are coupled through the hop count $\delta$ in Equation~\ref{eq:rendezvous}.
\sys satisfies these constraints in one of two flows, which trade compile time against schedule quality.
The space-first flow decides placement and then scheduling, whereas the cluster flow decides tensor placement and scheduling together in an ILP.

\begin{figure}[t]
\centering
\includegraphics[width=\columnwidth]{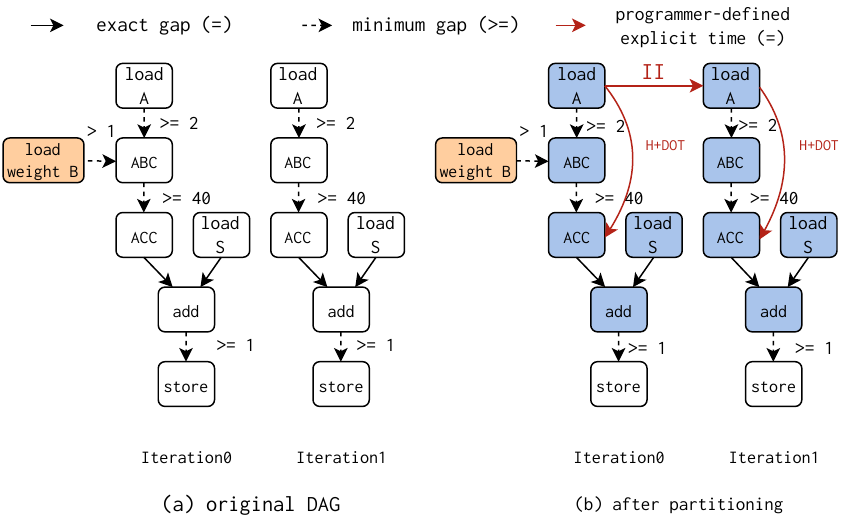}
\caption{Cluster partitioning of the row loop of \code{tile} (Figure~\ref{fig:motivation-tauvm}).
Explicit times (red) merge the blue operations into one cluster.}
\label{fig:clusters}
\end{figure}

\subsection{Space-First Scheduling}
\label{sec:greedy}

In a tensor kernel, regular data and computation patterns allow simple rules to assign every tensor to memory slices and words and each operation to a matching functional slice without search.
Fixing these slice locations turns every hop count $\delta$ in Equation~\ref{eq:rendezvous} into a known constant and decouples scheduling from placement.
To exploit this property for fast compilation, the space-first flow runs rule-based placement first and then schedules operations in cycle order with a greedy list scheduler~\cite{gibbons1986scheduling}, prioritizing operations by their critical-path latency to the end of the kernel.

For each loop, the scheduler builds a single-iteration schedule at either the programmer-specified initiation interval or the smallest interval at which no two iterations collide on a slice.
Repeating this schedule across the loop bounds advances the timeline to the end of the last iteration, so the next loop or operation starts only after the loop drains.
The scheduler requires every operation pinned with \code{.at} to issue at its written cycle and reports the source line of any mismatch.
While the space-first flow is fast and deterministic and serves as our baseline, separating placement from loop-by-loop scheduling limits schedule quality.
Because hop counts $\delta$ are fixed upfront, the list scheduler cannot relocate slices to align operand arrival cycles, which forces longer initiation intervals, while draining each loop prevents cross-loop overlap.

\subsection{Time-Guided Cluster Scheduling}
\label{sec:ilp}

Because placement and scheduling constrain each other, deciding both together finds schedules that no fixed placement allows, at the cost of a search space that grows exponentially in the number of operations.
In the cluster flow, \sys formulates this joint search as an ILP that decides which memory slices hold each tensor buffer and when every operation issues.
To keep the search tractable, as previewed in \S\ref{sec:examples-attention}, \sys uses stream dependences and explicit times to merge connected operations into \emph{timing clusters}, whose operations shift in time together (Figure~\ref{fig:clusters}b).

\noindent\textbf{Placement.}
Most of the spatial freedom lies in the memory slices, which far outnumber the vector, switch, and matrix slices.
How tensors are laid out across memory slices matters most for performance, because it sets the hop count $\delta$ that each value travels along SR tracks to reach its consumer.
The ILP therefore keeps the functional slices from the rule-based placement of the space-first flow and places only the tensor buffers.
A tensor buffer occupies consecutive memory slices starting from a single base slice, and each memory access runs on the slice that holds the part of the tensor it addresses.
The ILP thus carries one slice decision per buffer rather than one per access.

\noindent\textbf{Dependences.}
How a dependence constrains the schedule depends on the storage or communication medium it traverses (Figure~\ref{fig:clusters}a).
A dependence through memory banks or matrix weight planes is an inequality, because storage holds a value across cycles until another write overwrites it.
The consumer may therefore read the value at any cycle after the write commits.
In contrast, a stream carries each value from its producer to its consumer without waiting, so their issue cycles must obey the rendezvous equality of Equation~\ref{eq:rendezvous}.
This equality is where placement and scheduling constrain each other.
Choosing the two slices fixes the hop count $\delta$ and therefore the cycle gap between the two operations.
Conversely, once the two issue cycles are fixed, Equation~\ref{eq:rendezvous} fixes the hop count $\delta$ between their slices.

\noindent\textbf{Explicit times.}
Explicit times also state equalities between issue cycles (red arrows in Figure~\ref{fig:clusters}b).
A time map fixes the issue cycle of the operation that initiates each iteration, and a pin (\code{.at}) fixes the issue cycle of the pinned operation.
Every time map and pin therefore fixes the exact cycle gap between an operation and its root.

\noindent\textbf{Cluster partitioning.}
Given the operation DAG annotated with dependence bounds and explicit times, the issue cycles of its operations are not independent, because every exact gap (\code{=}) fixes the cycle difference between two operations before the solver runs.
To presolve these equalities, \sys partitions the DAG along all exact-gap edges, from both stream dependences and explicit times, into connected components (\emph{timing clusters}).
Because operations in the same cluster have fixed cycle offsets from one another, \sys substitutes out their individual issue cycles along a spanning tree, which leaves at most one unknown start cycle per cluster and none at all for the cluster rooted at \code{tau.T}.
Their slice placements are not independent either.
When two stream-connected operations fall into the same cluster, the cluster's start cycle cancels in Equation~\ref{eq:rendezvous} and fixes the hop count $\delta$ between them, so placing one slice directly determines the other at distance $\delta$.

Figure~\ref{fig:clusters} illustrates this presolve on the row loop of \code{tile} from \S\ref{sec:examples} (Figure~\ref{fig:motivation-tauvm}).
In the original DAG (Figure~\ref{fig:clusters}a), machine latencies give each edge either an exact gap (solid) or a minimum gap (dashed).
Because \code{add} takes two stream operands that must reach \code{VXM} in the same cycle, \code{ACC} fixes when both \code{add} and \code{load S} issue and joins the three operations with exact gaps within each iteration.
In contrast, \code{ABC} reads its weights from a weight plane and takes only \code{load A} as a stream input, while the remaining edges pass through the matrix accumulator or memory and therefore impose only minimum gaps on \code{load A}, \code{load weight B}, and \code{store}.
In Figure~\ref{fig:clusters}b, explicit times (red) add the missing exact gaps.
The time map of the row domain links \code{load A} across iterations at interval $\mathit{II}$, and the \code{.at} on the accumulation fixes \code{ACC} at $H + \mathit{DOT}$ cycles after \code{load A}.
Because $H$ is the hop count from \code{load A} to \code{ABC}, \code{ABC} has no slack between \code{load A} and \code{ACC}.
Since \code{load A} initiates each iteration from \code{tau.T}, partitioning merges all blue operations across iterations into a single cluster with known issue cycles, so the solver schedules only \code{load weight B} and \code{store}.
From the fixed cycle gaps inside this blue cluster, Equation~\ref{eq:rendezvous} also determines the hop counts from \code{load A} to \code{ABC} ($H$ hops) and from \code{load S} to \code{add}, so the memory slices of $A$ and $S$ follow directly from the matrix and vector slices.
The pin on \code{exp\_f32} in \S\ref{sec:examples-attention} forms its cluster the same way.

\noindent\textbf{Resources.}
While cluster partitioning fixes cycle offsets within each cluster, operations in different clusters can still share a slice and must not collide on the same cycle or unit occupancy.
Rather than unrolling a loop into one variable per iteration, the ILP keeps a single start-cycle variable for each rolled operation and forbids start-cycle differences at which two operations on the same slice would collide~\cite{rau1994modulo}.
For example, if operations $A$ and $B$ each run three times at a four-cycle interval ($a, a+4, a+8$ and $b, b+4, b+8$), they collide on a shared slice whenever $a - b \in \{-8, -4, 0, 4, 8\}$.
Before solving, \sys precomputes this forbidden set for every pair of operations that may share a slice.

\noindent\textbf{Objective and diagnostics.}
The ILP objective minimizes the kernel makespan, defined as the cycle by which every operation completes.
\sys solves the remaining slice and issue-cycle variables with CP-SAT~\cite{perron2023cpsat} under a fixed time budget.
When no valid schedule satisfies the explicit times, \sys reports the violated pins or the stream dependence whose rendezvous Equation~\ref{eq:rendezvous} cannot satisfy under any placement.

%% file: sections/implementation.tex
\section{Implementation}
\label{sec:implementation}

We implement \sys in about 44K lines of Python and C++ on top of MLIR~\cite{lattner2021mlir}.
A Python front end lowers kernels into our MLIR compiler pipeline, which carries symbolic timing attributes through call composition, affine time validation, and scheduling.
We implement the ILP of the cluster flow on top of the CP-SAT solver in OR-Tools~\cite{perron2023cpsat}.
Once slices and issue cycles are assigned, backend passes allocate inter-slice SR tracks, partition the program into one timestamped instruction stream per slice, and assemble the streams into a binary.

Finally, we implement a cycle-accurate software simulator and a SystemVerilog RTL model of the machine's control logic.
The simulator executes compiled binaries cycle by cycle across all slices, computes every data value, and traps on any rendezvous or resource violation, while the RTL model validates the simulator's cycle timing.

%% file: sections/evaluation.tex
\section{Evaluation}
\label{sec:evaluation}

Our evaluation answers five questions that progress from untimed compilation to explicit timing in \sys.
(1) How much does deciding placement and issue cycles together reduce schedule latency over space-first scheduling on untimed kernels (\S\ref{sec:eval-quality})?
(2) Does fixing placement upfront still lose optimality under an exact scheduler, and what dominates the joint search cost (\S\ref{sec:eval-ablation})?
(3) Which explicit times accelerate joint search without increasing schedule latency (\S\ref{sec:eval-written})?
(4) How well do explicit time pins scale joint search as kernel size grows (\S\ref{sec:eval-scale})?
(5) How do parameterized time maps let the programmer explore schedule trade-offs without rewriting the kernel (\S\ref{sec:case-study})?

\subsection{Evaluation Setup}
\label{sec:eval-setup}

\noindent\textbf{Metrics.}
We report kernel latency in execution cycles and compile time in wall-clock seconds.
On a statically scheduled machine, the binary fixes the slice and issue cycle of every instruction and each opcode has a fixed, data-independent latency,
so a single run determines a kernel's exact cycle count and small problem sizes suffice to validate the simulator against RTL.
We measure all kernels on our cycle-accurate simulator,
whose instruction latencies and issue intervals come from our 93-slice SystemVerilog RTL and match the published TSP parameters~\cite{abts2020thinkfast}.
Since full RTL simulation in Verilator is too slow for larger problem sizes,
we run the smallest instance of each kernel and one ResNet residual block in Verilator under both flows
and confirm that the RTL matches the simulator's cycle count on every binary without timing or resource violations.

\noindent\textbf{Configurations.}
We compare the two scheduling flows of \sys.
The baseline \emph{space-first flow} fixes slice placement first and runs a greedy list scheduler (\S\ref{sec:greedy}), whereas the \emph{cluster flow} runs joint search, solving the ILP of \S\ref{sec:ilp} with CP-SAT~\cite{perron2023cpsat} under a 60-second budget per solve.
The case study raises this budget to 100 seconds, and to 600 seconds for some pairs near its front.

\noindent\textbf{Workloads.}
We evaluate ten kernels spanning BLAS~\cite{lawson1979blas}, PolyBench~\cite{pouchet2012polybench}, ResNet~\cite{he2016resnet}, and Transformer layers~\cite{vaswani2017attention,zhang2019rmsnorm} (Table~\ref{tab:eval-kernels}).
Each kernel is evaluated at three problem sizes (30 instances in total) across three DSL versions.
The \emph{untimed} version writes no explicit times, the \emph{time maps} version replaces \code{range} loops with \code{tau.domain} and its time map, and the \emph{pins} version adds one to three lines of \code{.at}.
Sections~\ref{sec:eval-quality} and~\ref{sec:eval-ablation} compile the \emph{untimed} version in both flows to isolate the effect of joint placement and scheduling.
Section~\ref{sec:eval-written} then compares all three versions, while Sections~\ref{sec:eval-scale} and~\ref{sec:case-study} use the timed versions to scale joint search and explore schedule parameters.

\begin{table}[t]
  \centering
  \small
  \setlength{\tabcolsep}{3.5pt}
  \caption{The ten kernels, the slices each uses besides memory, and the three sizes of each.}
  \label{tab:eval-kernels}
  \begin{tabular}{@{}llll@{}}
    \toprule
    Kernel & Source & Slices & Sizes \\
    \midrule
    AXPY & BLAS 1 & VXM & 4, 8, 16 vectors \\
    GEMV & BLAS 2 & MXM & 4, 8, 16 vectors \\
    GEMM & BLAS 3 & MXM, VXM & 2, 3, 4 blocks of $K$ \\
    Jacobi-2D & PolyBench & SXM, VXM & 4, 8, 16 rows \\
    Conv2D $3{\times}3$ & ResNet & MXM, VXM & 4, 8, 16 rows \\
    MaxPool $3{\times}3$/2 & ResNet & VXM & 4, 8, 16 positions \\
    Softmax & Transformer & MXM, VXM & 4, 8, 16 rows \\
    RMSNorm & Transformer & MXM, VXM & 4, 8, 16 rows \\
    MLP, 2 layers & Transformer & MXM, VXM & 4, 8, 16 vectors \\
    Horner & activation & VXM & degree 4, 8, 16 \\
    \bottomrule
  \end{tabular}
\end{table}

\begin{figure}[t]
  \centering
  \includegraphics[width=\columnwidth]{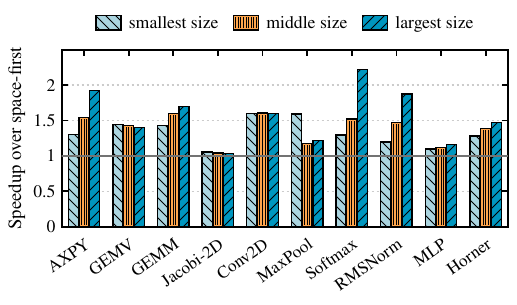}
  \caption{Latency speedup of the cluster flow over the space-first flow across the three problem sizes of Table~\ref{tab:eval-kernels}.}
  \label{fig:eval-quality}
\end{figure}

\subsection{Deciding Placement and Cycles Together}
\label{sec:eval-quality}

On a statically scheduled machine, placement and scheduling are coupled through stream hop counts (Equation~\ref{eq:rendezvous}).
We first test whether deciding both together is necessary for schedule quality by comparing the cluster flow against the space-first flow across all 30 untimed kernel instances.
As shown in Figure~\ref{fig:eval-quality}, the cluster flow finds a lower-latency schedule than the space-first flow on every instance and achieves a $1.40\times$ geometric-mean speedup (ranging from $1.04\times$ to $2.22\times$).
Three repeated compiles of each instance produce identical latency.

Two mechanisms explain the gain.
First, choosing memory placement together with issue cycles shortens loop initiation intervals.
Because the space-first flow fixes tensor placements before scheduling, operands streaming from different memory slices incur mismatched hop counts $\delta$, which forces the greedy list scheduler to pad the initiation interval to avoid collisions.
By placing memory buffers to align operand arrival cycles, the cluster flow reduces the initiation interval at the largest problem size from 19 cycles to 1 on Softmax, from 12 to 1 on RMSNorm, and from 4 to 1 on AXPY, so their speedups grow with the number of rows or vectors.
Second, joint search overlaps consecutive loops.
On GEMM and Conv2D, both flows achieve a 1-cycle initiation interval within each loop, but the space-first flow drains each loop completely before starting the next.
In contrast, the cluster flow schedules all loops together and overlaps each weight install with the tail of the rows streaming through the previous weight plane, which reduces the interval between the four weight installs of 4-block GEMM from 170--178 cycles to 84 cycles.

Across all ten kernels, the speedup of joint search grows with problem size wherever the cycles it saves recur in every iteration or block, as in AXPY, Softmax, RMSNorm, GEMM, and Horner.
Conversely, the speedups of GEMV, MLP, and Jacobi-2D stay nearly flat because fixed setup or serialized \code{SXM} shifts dominate their latency, while MaxPool peaks at the smallest size where its short loop allows tighter output spacing.

\begin{figure}[t]
  \centering
  \includegraphics[width=\columnwidth]{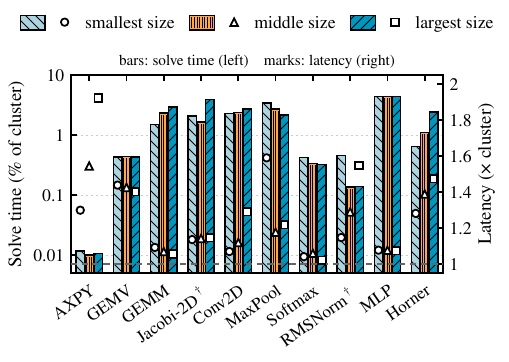}
  \caption{Solve time (bars, left axis, percentage) and latency (marks, right axis, ratio) of the fixed-placement flow relative to the untimed cluster flow.
    Bar colors and marker shapes denote problem size, and the dashed line marks equal latency.
    $\dagger$~marks kernels whose optimal latency remains unproven for at least one size.
    AXPY matches the optimal untimed reference latencies.}
  \label{fig:eval-ablation}
\end{figure}

\subsection{Placement Ablation}
\label{sec:eval-ablation}

Because space-first scheduling both fixes placement upfront and uses a greedy list scheduler, \S\ref{sec:eval-quality} alone does not show whether fixing placement still loses optimality under an exact scheduler.
To isolate the effect of fixed placement, we evaluate a \emph{fixed-placement flow} that solves the same untimed ILP while holding every buffer at the slice chosen by the space-first flow's rule-based placement.
As shown in Figure~\ref{fig:eval-ablation}, even though CP-SAT proves every fixed-placement schedule optimal for its given placement, fixed placement still increases latency across all 30 instances by $1.24\times$ in geometric mean (ranging from $1.02\times$ to $1.92\times$) compared to the cluster flow, which factors the $1.40\times$ speedup of \S\ref{sec:eval-quality} into $1.13\times$ from the ILP scheduler and $1.24\times$ from joint placement.
While the fixed-placement ILP still overlaps consecutive loops, CP-SAT proves that the 1-cycle initiation intervals reached by the cluster flow are infeasible under the rule-based placement on 19 of the 30 instances, so the minimum interval at the largest size rises to 2 cycles on Softmax, 8 on RMSNorm, and 4 on AXPY.
At the same time, fixed placement cuts solve time to at most 2.4 seconds per instance, or 0.01\% to 4.4\% of the cluster flow's solve time, which shows that searching memory slice placements accounts for over 95\% of the solve time.

\begin{figure}[t]
  \centering
  \includegraphics[width=\columnwidth]{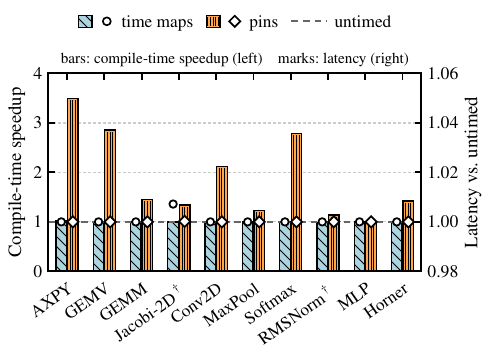}
  \caption{Compile-time speedup and relative schedule latency over the untimed kernel (geometric mean across the three sizes; $\dagger$~marks a kernel where a solve reaches the 60\,s budget).}
  \label{fig:eval-written}
\end{figure}

\subsection{Choosing What to Pin}
\label{sec:eval-written}

While joint search is necessary for schedule quality, searching memory slice placements and issue cycles together is costly.
To accelerate this search without raising latency, a \sys program can specify either loop initiation intervals via time maps or operation cycle offsets via explicit time pins.
Using the machine latencies exposed by \sys, our kernels each pin one to three recurring pipeline offsets with known optimal values,
such as the accumulation after a matrix multiply or the interval between weight installs.

Figure~\ref{fig:eval-written} compares the \emph{time maps} and \emph{pins} versions against the \emph{untimed} version (median compile time of three runs).
Explicit time pins cut compile time by $1.72\times$ in geometric mean and by up to $3.50\times$ on AXPY, while matching the untimed latency on 28 of the 30 instances and staying within one cycle of it on 8- and 16-row Jacobi-2D.
Time maps alone, by contrast, leave compile time unchanged ($1.00\times$ geometric mean), because they fix only the initiation interval, which the solver already finds in its first search round.
They also raise the latency of Jacobi-2D to $1.01\times$, because its time map starts the sum loop only after every shift loop finishes.

Pins shorten the search for memory slice placements, which dominates solve time.
Pinning the issue cycles at both ends of a stream dependence fixes the hop count $\delta$ between their slices through Equation~\ref{eq:rendezvous}, which prunes candidate placements.
The compile-time benefit therefore depends on how much placement search the pinned timing relation removes.
Consequently, kernels with more irregular data layouts, such as AXPY, Softmax, and GEMV, gain the largest compile-time speedups.
Because these layouts create many candidate placements for the untimed solver to compare, fixing their pipeline gaps with pins cuts compile time by $2.79\times$ to $3.50\times$.
In contrast, MLP and GEMM carry the same weight-to-accumulation pin as GEMV, but each phase repeats a uniform data layout, which yields smaller compile-time speedups of $1.00\times$ and $1.45\times$.
Beyond cutting compile time at these sizes, pruning candidate placements also changes how the search scales with kernel size.

\subsection{Scaling Joint Search}
\label{sec:eval-scale}

\begin{figure}[t]
  \centering
  \includegraphics[width=\columnwidth]{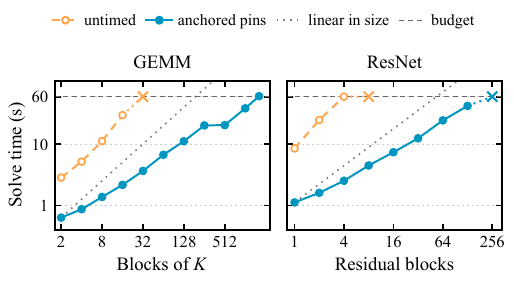}
  \caption{With pins, solve time grows more slowly than the kernel.
    The plot excludes the time to build the ILP.
    Circles mark sizes with a schedule.
    Crosses mark sizes with no schedule within the budget.}
  \label{fig:eval-scale}
\end{figure}

To test whether pins let the cluster flow scale, we grow two representative workloads far past the sizes of Table~\ref{tab:eval-kernels}.
First, we scale 16-row GEMM from 2 to 1,636 blocks of $K$, where each block adds a weight-install phase followed by a row-streaming phase.
Second, we stack 8-row ResNet residual blocks~\cite{he2016resnet}, each fusing two $3{\times}3$ convolutions and a skip connection.
In both, all blocks are fused into one kernel whose weights and intermediate buffers must fit on the chip together, so the compiler must place and schedule all phases globally.

We compile each workload both \emph{untimed} and with \emph{pins}, plus one pin between consecutive convolutions in ResNet, while the solver still places every buffer and weight set and schedules all unpinned operations.
Figure~\ref{fig:eval-scale} plots the total solve time per compile (averaged over three compiles) as kernel size grows.
We plot solve time rather than compile time because building the ILP takes the same time in both versions wherever both find a schedule.
The untimed solver stops scaling early.
Untimed GEMM solve time grows by $10.5\times$ from 2 to 16 blocks and finds no schedule within the budget at 32 blocks, while untimed ResNet reaches the budget at 4 residual blocks and finds no schedule at 8.

With pins, by contrast, the solver finds schedules at far larger sizes, and its solve time grows sub-linearly with kernel size.
As GEMM grows $818\times$ (from 2 to 1,636 blocks), solve time grows by only $97\times$ (from 0.63 to 61 seconds, staying near 20 seconds between 256 and 512 blocks), and as ResNet grows $128\times$ (from 1 to 128 residual blocks), solve time grows by only $38\times$ (from 1.1 to 42 seconds, with no schedule at 256 blocks).
Every completed solve proves its schedule optimal under its pins.
Across the sizes where the untimed solver succeeds, pins reduce solve time by $4.5\times$ to $15.5\times$ with zero latency penalty and match the exact untimed growth rate of 96 cycles per $K$-block in GEMM and 648 cycles per residual block in ResNet.
On 8-block ResNet, the single pin between consecutive convolutions cuts solve time by a further $3.3\times$ over intra-convolution pins alone (from 14.9 to 4.5 seconds).

Once timing clusters tame the search itself, scaling to these sizes exposes the next largest cost in compile time, which is building the ILP before the solver runs.
Building the ILP involves no search, as it only precomputes the forbidden start-cycle differences across operation pairs that may share a slice.
This step takes under one second up to 64 GEMM blocks and 8 residual blocks, and reaches 182 and 194 seconds at the largest sizes, $3.0\times$ and $4.6\times$ as long as the solve.
Overall, a few pins in \sys let the solver prove schedules optimal under those pins at 1,636 GEMM blocks and 128 residual blocks,
far beyond the sizes the untimed solver reaches.

\subsection{Exploring Schedules by Parameters}
\label{sec:case-study}

\begin{figure}[t]
  \centering
  \includegraphics[width=\columnwidth]{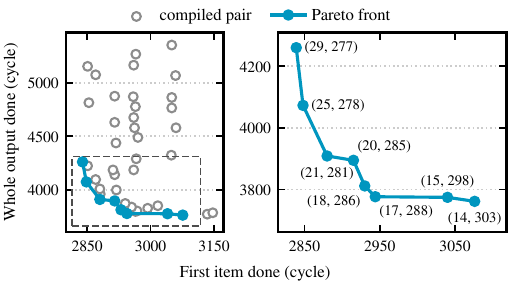}
  \caption{Exploring the trade-off between first-item latency and whole-output time on a single kernel.
    Each circle is a compiled schedule for a parameter pair $(\mathit{II}, \mathit{STEP})$ at $N = 50$ and $K = 10$, and the solid line connects the Pareto front of the compiled pairs.
    The right panel enlarges the front inside the dashed region and labels each front schedule with its $(\mathit{II}, \mathit{STEP})$ values.}
  \label{fig:case-study}
\end{figure}

Although time maps alone do not accelerate solver search,
parameterizing their strides lets the programmer explore schedule trade-offs on the same kernel, as in the GEMM schedules of Figure~\ref{fig:schedules}.
We demonstrate this on a flow-matching action head that refines a chunk of $N = 50$ robot actions over $K = 10$ sequential Euler steps, following $\pi_0$~\cite{black2025pi0}.
Reactive control minimizes first-item latency so the robot acts on the earliest action~\cite{hoeg2024sdp,jiang2025sfp},
whereas chunked execution minimizes whole-output time across the full chunk~\cite{black2025rtc}.
Because the kernel omits cross-item attention, its 50 items execute independently,
while within each item step $k+1$ depends on the two matrix projections and Euler update of step $k$.

A single 2D time map, \code{tau.T + a*II + k*STEP}, schedules step $k$ of item $a$,
spacing successive items by $\mathit{II}$ cycles and successive steps of the same item by $\mathit{STEP}$ cycles.
Only these two stride parameters change across compiles.
Reducing $\mathit{STEP}$ lowers first-item latency because the final step of the first item starts $(K-1) \cdot \mathit{STEP}$ cycles after its first,
whereas reducing $\mathit{II}$ lowers whole-output time because the last item starts $(N-1) \cdot \mathit{II}$ cycles after the first.

We first screen 891 parameter pairs across $\mathit{II}$ from 16 to 48 cycles and $\mathit{STEP}$ from 274 to 300 cycles.
Before invoking the solver, affine time validation rejects 407 of these pairs within about a second each,
because their strides assign two matrix operations to the same slice in the same cycle.
A compile that finds a schedule takes 32 to 518 seconds (median 68),
so we solve 84 of the 484 accepted pairs on a coarse grid of $\mathit{II}$ from 20 to 48 with six $\mathit{STEP}$ values
and sample 30 more pairs around the grid's front ($\mathit{II}$ down to 13, $\mathit{STEP}$ up to 311).
Of these 114 pairs, 44 yield a schedule and 70 reach the budget,
mostly at tighter strides where a larger budget could expand the front further.

Eight of the 44 schedules form the Pareto front among compiled pairs in Figure~\ref{fig:case-study}.
Along the front, $\mathit{II}$ falls from 29 to 14 cycles while $\mathit{STEP}$ rises from 277 to 303.
Reactive control selects $(29, 277)$, which lowers first-item latency by 237 cycles compared to $(14, 303)$ (2839 versus 3076 cycles),
whereas chunked execution selects $(14, 303)$, which reduces whole-output time by 498 cycles (3762 versus 4260 cycles).
Across all 44 schedules, whole-output time exceeds first-item latency by exactly $49 \cdot \mathit{II}$ cycles,
and first-item latency falls between $9 \cdot \mathit{STEP} + 346$ and $9 \cdot \mathit{STEP} + 360$ cycles.
The programmer can thus predict both latencies from $(\mathit{II}, \mathit{STEP})$ to within 14 cycles before compiling, while the compiler determines which pairs are realizable.
Adapting the kernel between control regimes requires only changing two parameter values rather than rewriting the kernel.

%% file: sections/related-work.tex
\section{Related Work}
\label{sec:related}

\noindent\textbf{Tile-level kernel languages.}
Triton~\cite{tillet2019triton}, TileLang~\cite{wang2026tilelang}, CuTile~\cite{nvidia2025cutile}, Tilus~\cite{ding2025tilus}, Helion~\cite{pytorch2025helion}, and Gluon~\cite{openai2025gluon} let a programmer express block-level computation over tiles while the compiler chooses layouts, warps, and pipelines.
\sys retains this tile-level programming model of loads, stores, matrix multiplies, and epilogues, but adds first-class cycle times so the compiler can statically schedule targets that lack dynamic hardware synchronization.
While Exo~\cite{ikarashi2022exo} also exposes scheduling to the programmer through explicit rewrites, its schedules specify operation orderings rather than issue cycles.

\noindent\textbf{Timing in hardware languages.}
High-level hardware design languages such as Spatial~\cite{koeplinger2018spatial}, Dahlia~\cite{nigam2020dahlia}, and Aetherling~\cite{durst2020aetherling} raise accelerator design above the register-transfer level while leaving cycle timing to the compiler.
Closer to our philosophy, Filament~\cite{nigam2023filament} and Lilac~\cite{nigam2026lilac} embed cycle intervals into the type system to verify hardware pipelines statically.
However, because these languages synthesize custom circuits, the designer can insert registers or buffers whenever a path needs delay.
In contrast, \sys targets a fixed accelerator with predetermined latencies, so it satisfies explicit times by scheduling operations jointly with tensor placement across memory slices.

\noindent\textbf{Compilers for spatial accelerators.}
TileLoom~\cite{li2026tileloom} similarly compiles tile programs to spatial accelerators by planning tile placement and data movement, but its schedule stops at spatial assignment and execution order.
CGRA compilers assign both spatial units and modulo cycles~\cite{mei2003dresc,rau1994modulo,liu2019cgrasurvey}, often using integer programming~\cite{nowatzki2013constraint,huang2021cosa}.
However, because CGRAs buffer early operands in local registers, data dependences only bound consumer issue cycles from below.
On our target, stream operands obey the exact rendezvous equality of Equation~\ref{eq:rendezvous}, which \sys combines with programmer-written explicit times to merge operations into timing clusters before solving.

\noindent\textbf{Polyhedral scheduling.}
Polyhedral compilers represent loop schedules as affine functions of the iteration indices and derive them automatically over data dependences~\cite{feautrier1992affine}.
\sys adopts this affine representation for its domain time maps, but lets the programmer write the schedule directly as cycle times while the compiler validates dependences algebraically.
Whereas traditional polyhedral schedules specify logical ordering for dynamically synchronized hardware, Clockwork~\cite{huff2021clockwork} interprets affine schedules as physical cycles to synthesize and size buffers for image-processing hardware.
In \sys, time maps likewise denote exact cycles, but because the target hardware is fixed, a valid schedule must satisfy Equation~\ref{eq:rendezvous} across existing slices and memory banks.

%% file: sections/conclusion.tex
\section{Conclusion}
\label{sec:conclusion}

\sys is, to our knowledge, the first kernel language and compiler for statically scheduled accelerators.
By making cycle time a first-class, parameterized language construct, \sys decouples a kernel's schedule from its code structure and enables whole-program schedule exploration without rewriting code.
Lowering explicit times into \sysir lets the compiler validate timing algebraically before search and merge connected operations into timing clusters during joint space-time scheduling.
Across our benchmarks, a few explicit time pins accelerate joint search and scale it to thousand-block kernels while matching untimed schedule latency within a single cycle.